# How lovebirds fly in crosswinds based on minimal visual information

Daniel Quinn[1,2]*, Daniel Kress[1], Eric Chang[1], Andrea Stein[1], Michal Wegrzynski[1], David Lentink[1]*

[1]Stanford University
[2]University of Virginia
*danquinn@virginia.edu and *dlentink@stanford.edu

Flying birds navigate effectively through crosswinds, even when wind speeds are as high as flight speeds. What information birds use to sense crosswinds and compensate is largely unknown. We found that lovebirds can navigate 45-degree crosswinds similarly well in forest, lake, and cave-like visual environments. They navigate effectively using only a dim point light source as a beacon, despite being diurnal and raised in captivity. To maintain their heading, the lovebirds turn their bodies into the wind mid-flight, while orienting their heads towards the goal with neck angles up to 30 degrees. We show how this wind compensation can be achieved using a combination of passive aerodynamics and active control informed by muscle proprioception, a sensory input previously thought to be unimportant in detecting wind.

To avoid drifting off course, birds sense crosswinds and compensate – both over long ranges, such as during inter-continental migration (*1*), and over short ranges, such as during treetop landings. Migrating birds navigate by fusing visual, olfactory, auditory, magnetic, and vestibular cues (*2*). Little is known about which of these senses help birds fly in crosswinds (*1*), because migration studies often cannot access sensory information (*3–6*). In near-ground short-range navigation, however, it is thought that the velocity of image patterns (optic flow) over the retina (*7*) enables birds to compensate for crosswinds (*1*). Accurate vision in flight is made possible by head stabilization, which reduces retinal blur (*11*), stabilizes image features (*12*), and defines a clear gravity vector for the vestibular system (*13*). Birds stabilize their heads using complex neck motions to negate wind perturbations (*8*), wing flapping (*9*), and flight maneuvers (*10*) – even if the body is fully inverted, as it can be for geese (*9*). How much visual information flying birds need to stabilize their heads and compensate for crosswinds remains unclear. Human pilots cannot fly safely at night without satellite and radar-based weather reports, the Global Positioning System, runway lighting, radio beacons, and guidance from air traffic controllers (*17*). Birds have none of this technology yet navigate even when visual cues are sparse. Frigate birds, for example, compensate for strong winds over open water, at times soaring in thick clouds (*15*), and specialist birds such as swiftlets and oilbirds fly in dark caves (*16*). What information birds need to compensate for crosswinds without becoming disoriented is an open question.

To determine what information birds need to navigate crosswinds, we manipulated the visual and wind environment in a 3.2 m long arena in which lovebirds (*Agapornis roseicollis*) flew between two perches (Fig. 1A). These generalist, diurnal, nonmigratory birds were raised in captivity, so they were naïve to crosswinds. We simulated three visual environments: a *cave* (our control case), black walls with a small dim light behind the goal perch to simulate a narrow cave exit; a *lake*, a horizontal contrast line to simulate the wide-field horizon of open water; and a *forest*, vertical stripes optimized for optic flow perception (Fig. 1B; Methods). The global illumination in the *cave* was only 0.2 lux, similar to a full moon (*18*). In contrast, the illumination in the *lake* and *forest* was 160 lux, similar to a closed canopy during a clear sunny day (*16*). We used movable wind generators to simulate three wind environments: *still*, no wind; *gust*, same-side crosswinds; and *shear*, opposing-side crosswinds (Fig. 1C). The wind speeds were comparable to the birds' flight speeds, causing effective wind angles up to 45°. To determine how the birds flew in the nine environmental permutations, we tracked 3D marker clusters on their head and body at 1000 Hz over 366 flights. The differences between the marker cluster attitudes quantify the neck's role in wind compensation (experimental details in Methods).

To our surprise, the lovebirds easily traversed the arena in all nine combinations of visual and wind environments. We only found small differences in the average ground speed between the visual conditions: lovebirds traverse the arena somewhat faster in the *forest* and somewhat slower in the *cave* (Fig. 1D). These findings agree with earlier observations that visual cues affect ground speed (*7*). Remarkably, the ground speed is not affected by wind environment; the birds reach the goal just as quickly in strong crosswinds as they do in still air (Fig. 1D). To accomplish this feat, the birds increase their airspeed by 44±18% (Fig. SF1), a strategy also seen in long-range flights of raptors (*19*) and fruit bats (*20*). Concurrently, the birds yaw their body towards the wind and their head towards the goal, regardless of the visual condition (Fig. 1E).

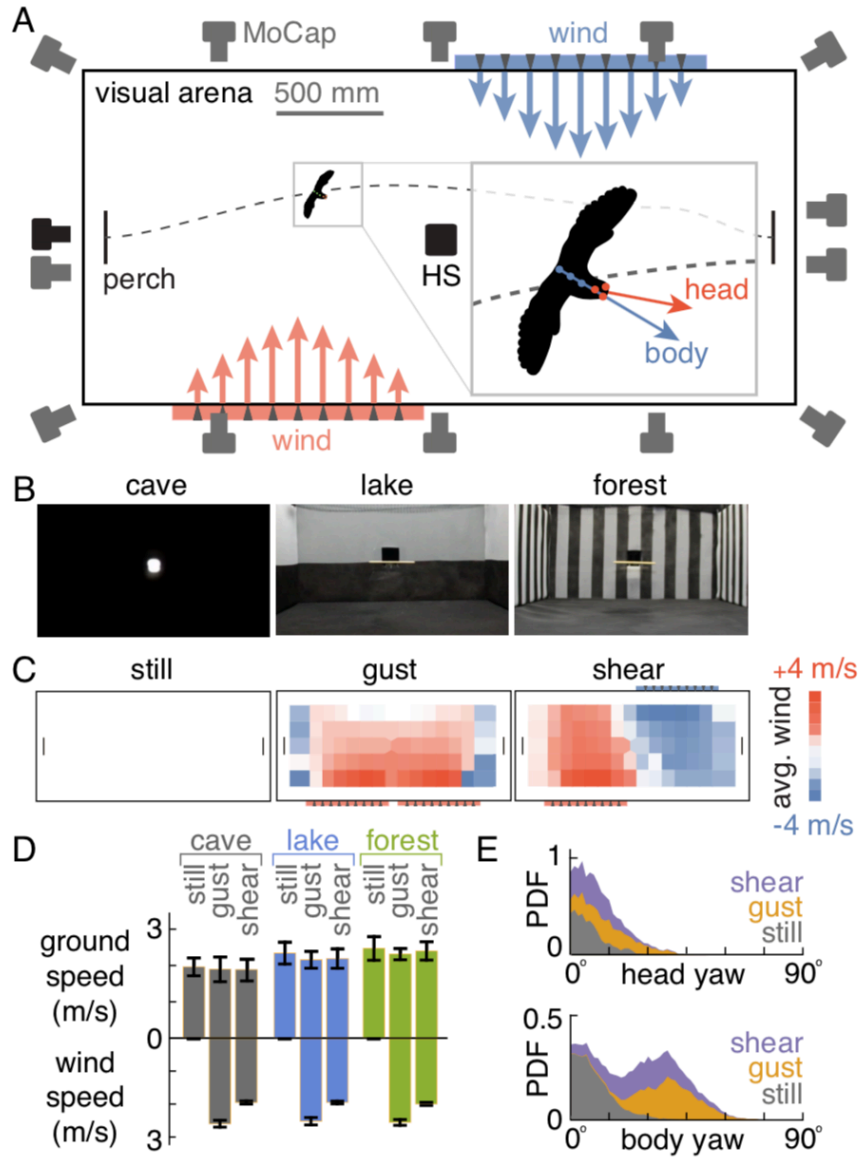

**Fig. 1. Lovebirds fly at constant ground speed towards a goal irrespective of crosswinds by turning their body into the wind.** (**A**) Flight arena. Using IR light, 13 cameras automatically tracked marker clusters on the body and head (1000 Hz), and 2 grayscale cameras recorded video (500 Hz). (**B**) Visual environments simulated a *cave* (uniform black), *lake* (horizontal stripe), and *forest* (vertical stripes). (**C**) The wind generators produced *still* (no wind), *gust* (side wind) and *shear* (opposing side wind) environments. (**D**) The ground speed of lovebirds is somewhat higher in visually richer environments, but not modified by wind condition. (**E**) Lovebirds orient their head towards the goal and their body into the wind across visual conditions (PDF = Probability Density Function; histograms are stacked).

The lovebirds use fast neck motions to keep their head oriented towards the goal, even when flying towards a dim light. We illustrate this ability with bird BB's first flight in the *cave-gust* environment (Fig. 2A-C). BB showed no patterned changes in behavior with more time in the arena, indicating that her ability is mostly innate. Her head is stabilized over roughly three timescales: body oscillations during each wingbeat, body wobbling over several wingbeats, and body reorientations into the wind over approximately ten wingbeats (Fig. 2C). Head orientation changes primarily via pitch, roll, and yaw saccades, three of which occur simultaneously in this sample flight. Lovebirds seem to make fewer, larger saccades in the *cave* environment (ESM Fig. SF5), probably because gaze slip is harder to detect. Pitch saccades are infrequent (ESM Fig. SF5), which helps the vestibular system sense the direction of gravity, according to a robot model of head stabilization (*21*). Across all flights, lovebirds use their necks to actively stabilize head yaw and roll relative to the body (Fig. 2D-F). Head roll and yaw are coupled (Fig. 2D-F), much like airplane roll and yaw can be during maneuvers (*22*). This coupling must rely on vestibular and proprioceptive cues, because the point light in the *cave* provides very little roll information. Head stabilization over a range of timescales is consistent with other vertebrates, where the neck's muscle tone passively dampens high frequency motions (*13, 23*) and the vestibulocollic and cervicocollic reflexes actively dampen low frequency motions (*24, 25*).

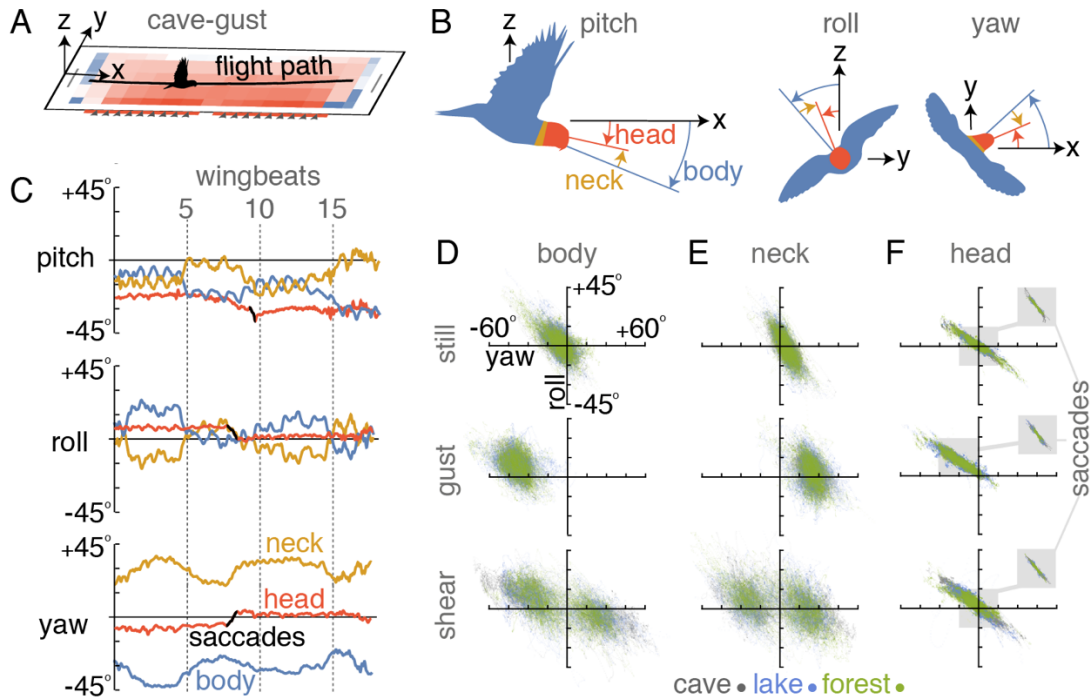

**Fig. 2. The neck isolates the head from body motion and orients it towards the goal based on minimal visual information.** (**A**) The first flight by bird BB in the cave-gust environment. We consider motions in the inertial arena frame (x, y, z). (**B**) The difference between head and body attitude determines neck pitch, roll, and yaw joint angles. Negative pitch denotes a "pitch up", chosen such that positive z is upward. (**C**) The neck isolates the head from body motion over multiple timescales. (**D**) Body yaw and roll are not strongly coupled (slope -0.7; $R^2$ 0.37), and yaw is offset in crosswinds. (**E**) The neck stabilizes the head by counteracting offsets in body angle. (**F**) Head yaw and roll are proportionally coupled (slope -0.6±0.3; $R^2$ 0.75) with offset depending on crosswind. Inset; 70% of the saccades simultaneously yaw and roll the head (slope -1.4±0.4; $R^2$ 0.98).

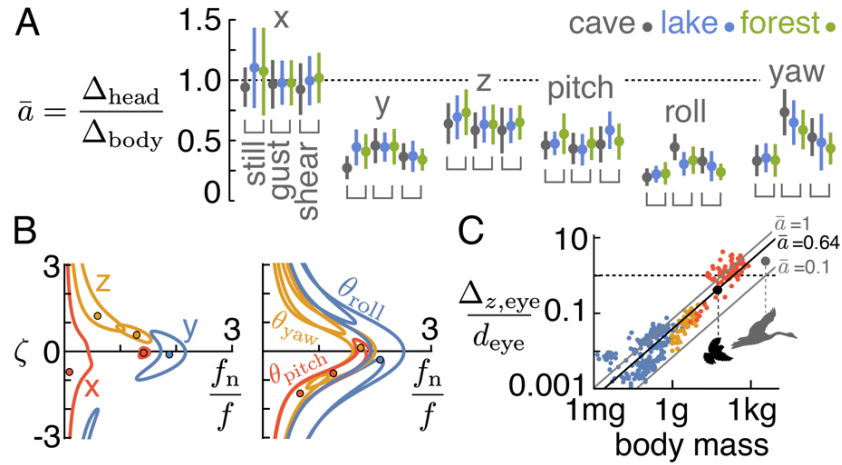

**Fig. 3. Lovebirds stabilize their head beat-by-beat in all directions except frontal.** (**A**) Whereas motion in the frontal (x) direction is not stabilized (gain $\bar{a} \approx 1$), the neck stabilizes lateral (y), vertical (z), pitch, roll and yaw motion. Head stabilization in each wind condition is similar across visual environments (underbars represent wind conditions). (**B**) A neck suspension model reveals ranges of natural frequency ratios ($f_n/f$) and damping coefficients ($\zeta$) corresponding to gains and phase lags observed in the head motion (circles, mean; contours, +/- $\sigma$). (**C**) Smaller flying animals can maintain vertical image jitter less than eye diameter ($\Delta_{z,\text{eye}}/d_{\text{eye}} < 1$) regardless of the head-body gain ($\bar{a}$). The avatars indicate lovebird and mute swan data; blue dots, insects; yellow, hummingbirds; red, birds (*32*). See ESM S1 for scaling details.

    The gain and phase of the neck reveal that high frequency linear head motions can be attenuated passively. By high-pass filtering the neck motions, we found that lovebirds attenuate lateral (y) and vertical (z), but not frontal (x) motions (Fig. 3A). The lateral and vertical residual head amplitudes are 22% +/- 6% and 43% +/- 10% of the eye diameter (5 mm), which reduces retinal image jitter (*9*). This reduction is probably not essential for frontal motion, where jitter will not substantially change motion parallax (*26*). Minimal frontal stabilization may be advantageous, because a force transmission ratio near 1 helps birds estimate distance by integrating acceleration (*27*). By applying a semi-passive neck suspension model, we found that the lovebird's S-shaped neck acts like a tuned anisotropic viscoelastic beam. In the frontal direction, the head is nearly in phase with the body (phase lag: -0.02 +/- 0.08 wingbeats), corresponding to either a strut-like stiff spring (high muscle tone), or a narrowly tuned spring (lower muscle tone) with almost no damping (Fig. 3B). The lateral and vertical directions (phase lag: -0.03 +/- 0.20 wingbeats and 0.18 +/- 0.09 wingbeats) correspond to under-damped and critically-damped springs. Motivated by the success of the suspension model for both lovebirds and whooper swans (*9*), we analyzed how vertical head attenuation scales with body mass. Using isometric scaling (ESM S1), we predict that insects and hummingbirds experience minor vertical image jitter, while larger birds experience major jitter that must be attenuated.

    Unlike linear head motion, angular head motion is attenuated more actively. The pitch, roll, and yaw residual amplitudes are similar: 2.5° +/- 0.6°; 1.9° +/- 0.7°; and 1.9°

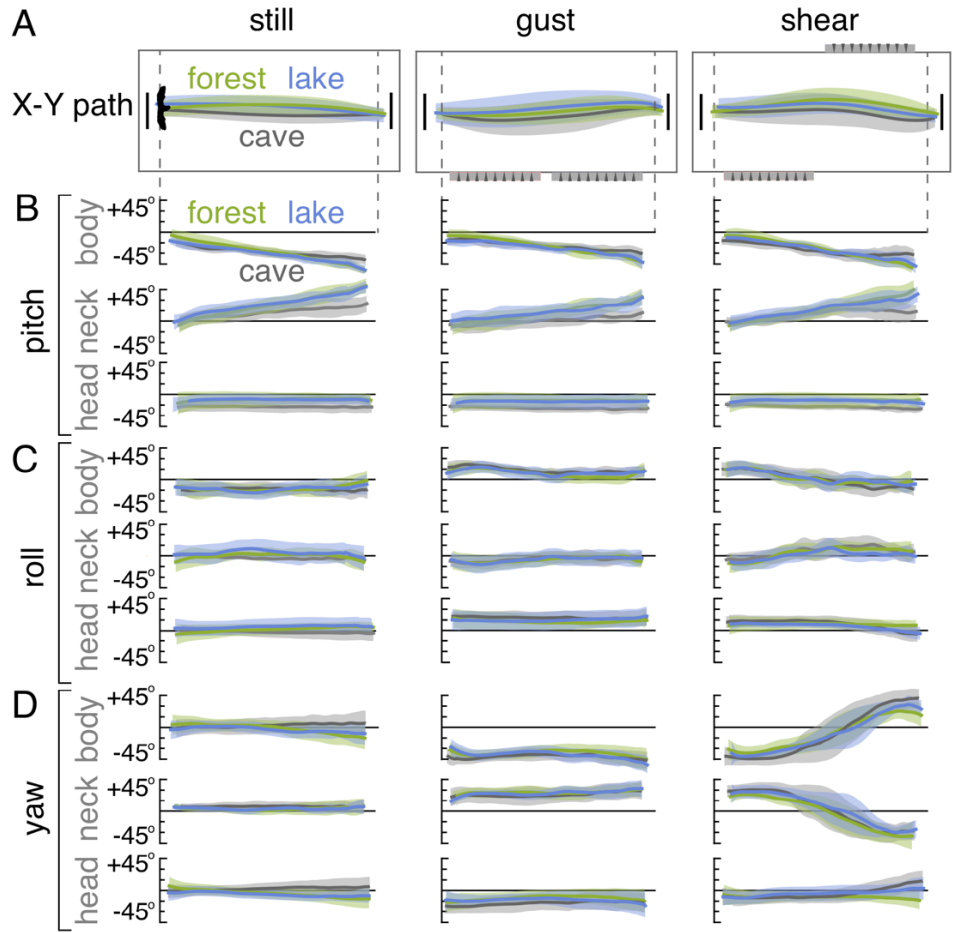

**Fig. 4. Lovebirds traverse complex wind environments similarly well in forest, lake, and cave visual environments by fixating on a goal.** (**A**) The average horizontal flight paths are similar across visual environments. (**B**, **C**, **D**) The pitch, roll, and yaw orientations of the head and body are similar across visual environments. Lovebirds pitch their body up gradually as they get closer to the landing perch while keeping their head level (B). In *gust* and *shear* conditions, lovebirds yaw their body into the wind while keeping their head fixated on the goal perch (D).

+/- 0.6°, and the relative reduction in roll is the most pronounced (Fig. 3A). When applying our semi-passive neck suspension model, the uncorrelated phase lags in head roll, pitch, and yaw (pitch, -0.20 +/- 0.33; roll, -0.06 +/- 0.39; yaw, 0.03 +/- 0.32) correspond to an envelope of torsional spring-damper coefficients (Fig. 3B). The prevalent negative damping ratios represent active motor-like muscle function. Combined with the spring-like properties of linear attenuation, these findings exhibit the known motor-, brake-, strut-, and spring-like functions of muscles (*28*). The maximum angular velocity of the head (residual amplitude $\times$ 2 pi $\times$ flapping frequency $\approx$ 250°/s), is larger than what small parrots can resolve at full resolution (visual acuity (*29*) $\times$ flicker fusion frequency (*30*) $\approx$ 0.1° $\times$ 70 Hz = 7°/s) (*11*). Lovebirds may therefore stabilize the image on their retina further via the vestibulo-ocular reflex (*13, 31*), though studies of pigeons suggest these attenuations may be limited to one sixth those of head stabilization (*31*). By stabilizing their head within a wingbeat, the lovebirds improve the accuracy of visual, proprioceptive, and vestibular cues that help them navigate over multiple wingbeats.

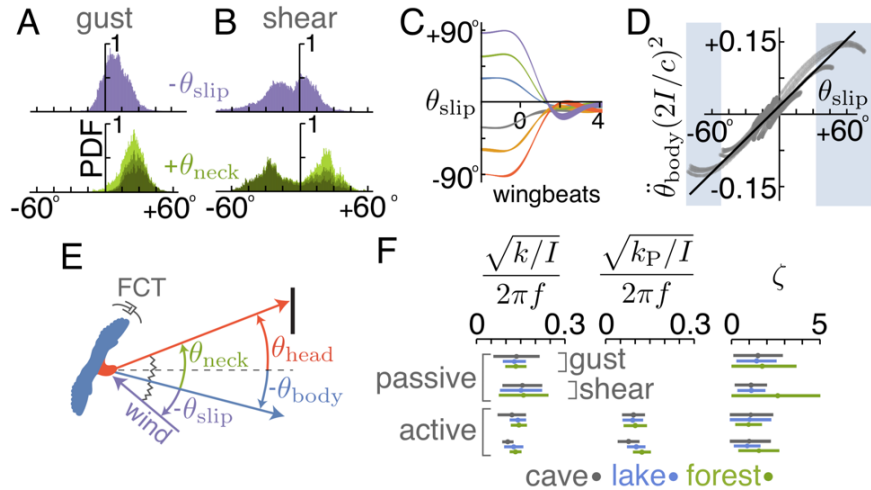

**Fig. 5. Using an inertia-spring-damper model for body yaw, we find lovebirds use passive and active control to reach the goal perch.** (**A**) Slip angles range up to 30° in the *gust* environment. Neck angles are sufficient for proprioceptive input (dark green, gaze aligns with perch; medium green, gaze within +/- 150% of the perch). (**B**) When landing in the *shear* environment (left side of plot), slip angles are larger and the head is fixated on the perch more often. (**C**) The slip angle passively goes to zero for an ornithopter, except for a small offset (≈8°) due to minor wing asymmetries. (**D**) The non-dimensional restoring torque on the ornithopter is proportional to slip angle over angles relevant to lovebirds (95% of lovebird data occurred between shaded boxes). Gray, tracking data; black, linear fit. (**E**) Body yaw is driven by a passive torque proportional to slip angle ($\theta_{\text{slip}}$) and an active torque proportional to neck angle ($\theta_{\text{neck}}$), and is dampened by Flapping Counter Torque (FCT). (**F**) The average corroborated coefficients in Eqn. 1 and 2 are similar across visual/wind environments; error bars show standard deviation.

Regardless of crosswinds, lovebirds compensate as well in the dark *cave* as they do in well-lit environments with a wide-field horizon (*lake*) or strong optic flow (*forest*) (Fig. 4). Across all conditions, the low-pass filtered flight paths (Fig. 4A) and body reorientations (Fig. 4B-D) are similar. The most pronounced reorientation is in yaw: the body orients roughly 45° into the crosswind and re-orients almost 90° midflight in the *shear* environment (Fig. 2D; 4D). The body also rolls into the wind, albeit over smaller angles (Fig. 2D; 4C). In all conditions, head pitch remains constant while the body pitches up in preparation for landing (Fig. 4B). The lovebirds combine strategies of general aviation pilots, who pitch the fuselage up and use either "crabbing" (yaw) or "wing-low" (roll) to compensate for strong crosswinds on final approach (*22*). Unlike airplanes, where both the fuselage and cockpit orient into the wind, lovebirds contort their necks (Fig. 5A-B) to fixate their heads on the goal perch (Fig. 1E). Motivated by the consistent yaw reorientations (Fig. 4D), we corroborated a yaw dynamics model to see if yaw could be controlled semi-passively.

Our model (Fig. 5E) shows that lovebirds can automatically direct their body into crosswinds, because flapping wings passively orient into the wind (Fig. 5C). To understand this passive reorientation, we tested a mechanical bird model—an ornithopter—and found that it passively reduced the slip angle (effective wind angle minus body angle) without needing a tail (Methods, Fig. 5C). The weathervane-like restoring torque on the flapping wings is proportional to the slip angle ($\theta_{\text{slip}}$) up to large angles (Fig. 5D). Flapping wings are known to automatically dampen yaw motion via 'Flapping Counter Torque', a torque proportional to body yaw velocity, $\dot{\theta}_{\text{body}}$ (*33*). These two passive aerodynamic torques sum together to determine the angular acceleration of the body yaw as follows:

$$\ddot{\theta}_{\text{body}} = (k/I)\theta_{\text{slip}} - (c/I)\dot{\theta}_{\text{body}}, \qquad (1)$$

where $I$ is the animal's moment of inertia about the vertical axis, $k$ is the aerodynamic restoring torque constant, and $c$ is the aerodynamic angular damper constant (details in Methods). By using the average corroborated $k/I$ and $c/I$ coefficient ratios, the model explains most of the body yaw dynamics in the *gust* and *shear* condition (*gust*, $R^2$ 0.87, RMS 12°; *shear*, $R^2$ 0.60, RMS 16°; measurement uncertainty ±2°). The residual error is appreciable because the passive model predicts a 0° equilibrium slip angle, whereas lovebirds maintain slip angles of 15° or more (Fig. 5A-B). Lovebirds must therefore actively control slip angle. We inferred that a proportional (P) controller may be sufficient for maintaining nonzero sideslip, because P controllers are known to produce similar offsets in their output. Our recordings show that the birds fixate their gaze on the goal (Figs. 1E; 4B-D) by actively contorting their neck up to 30° or more in pitch and yaw (Figs. 4B,D; 5B,C). These angles are large enough to give attitude information via muscle proprioception (*13*) according to data for humans (*34*). We also found that minor wing asymmetries result in nonzero equilibrium slip angles of the ornithopter (Fig. 5C). Therefore, we hypothesized that neck angle times a constant gain, $k_{\text{P}}$, could provide the control torque that the birds apply using wing asymmetry:

$$\ddot{\theta}_{\text{body}} = (k/I)\theta_{\text{slip}} + (k_{\text{P}}/I)\theta_{\text{neck}} - (c/I)\dot{\theta}_{\text{body}}. \qquad (2)$$

Indeed, adding the proportional controller enables the model to represent the data well (*gust*, $R^2$ 0.94, RMS 9°; *shear*: $R^2$ 0.79, RMS 13°). The goodness of fit especially improves in the *shear* case, where the 90° mid-flight yaw reorientations presumably require more active control. The fitted damping coefficients, $\zeta \equiv c/\sqrt{4(k + k_{\text{P}})I}$, cluster around 1 (Fig. 5F), which corresponds to the fastest possible crosswind response without oscillation. To prevent drift, the lovebirds could also use neck angle to modulate airspeed: we found that for each bird, lateral airspeed was directly proportional to neck angle ($R^2$ = 0.79, 0.77, 0.59 respectively; details in Methods).

Our corroborated model explains how lovebirds can navigate crosswinds based on muscle proprioception. This runs contrary to the established assumption that proprioception is unlikely to give wind information (*1*). By stabilizing both high and low frequency head motions, the lovebirds obtain a precise goal heading. By comparing that heading to body angle—which responds passively to the wind—the lovebirds can estimate the local wind angle. Wind may also be coarsely detected by filoplumes (*35*), which cause behavioral changes when wind is blown at the breast of a fixed bird (*36*), but neck proprioception offers comparatively rapid, high-fidelity, directional feedback, as it does for the cervicocollic reflex (*13*). The similar model coefficients across environments (Fig. 5F) show that this strategy works equally well in the dark. The rate at which the lovebirds can glean wind information from proprioception scales with the speed of the yaw response. According to our model, body yaw angle responds with a time constant $\tau = 2fI/c$ measured in wingbeats (ESM S3). This time constant generally decreases with body size—ranging from around ten wingbeats for insects to one wingbeat for large birds—showing that bigger animals could use proprioception to gather wind information over fewer wingbeats. The time constant is further modified by parameters such as air density, aspect ratio, and stroke amplitude within one order of magnitude (Fig. 6B).

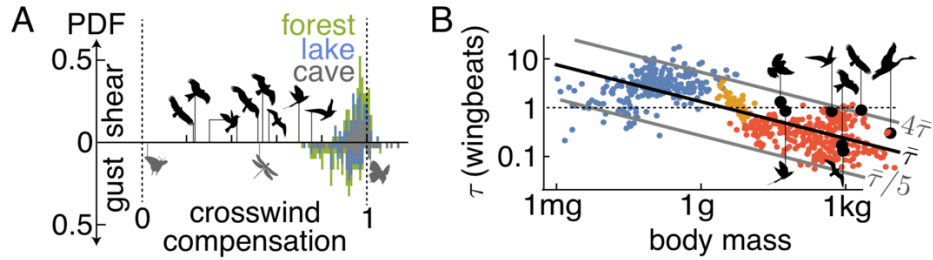

**Fig. 6. Lovebirds can use proprioceptive cues to effectively compensate for crosswinds over a wingbeat or less**. (**A**) Regardless of visual condition, lovebirds effectively compensate for crosswinds (0 = full drift; 1 = full compensation) compared to migratory species (table of avatar species in ESM Tab. ST1). (**B**) Gust information can be inferred from neck angle as body yaw responds to crosswinds, which occurs over fewer wingbeats for larger flying animals (table of avatar species in ESM Tab. ST2). Air density and non-dimensional ratios can change the yaw response time in wingbeats, $\tau$, up to an order of magnitude. Blue dots, insects; yellow, hummingbirds; red, birds (*32*). See ESM S4 for scaling details.

    Lovebirds appear to fuse proprioceptive, vestibular, and minimal visual cues to navigate over short ranges. Even a dim approximate point-source of light provides a sufficient goal heading. By stabilizing their head, the lovebirds improve the accuracy of their inertial horizon (*13*), their distance integration (*27*), and their estimates of wind *versus* goal direction via proprioceptive cues (Fig. 5A-B). Lovebirds surely fuse other cues as well, such as expanding (looming) visual cues (*37*), parallax (*38*), or spatial memory cues (*39*). However, given the predictive strength of our model (Eqn. 2), the most parsimonious explanation is that lovebirds navigate in the dark primarily by using the dim point light source as a reference for proprioceptive cues. By doing so, the lovebirds navigate through crosswinds without needing wide-field motion cues. This reliance on proprioception helps explain why pigeons fly poorly when a paper collar blocks neck motion (*10*). Optic flow is likely used to regulate ground speed (Fig 1D) and vertical position—as it is with budgies (*7*) and hummingbirds (*40*) in corridors of still air—but is apparently not essential for dynamic flight control. We infer from these findings that a wide-field horizon may be replaced by an inertial horizon, similar to how man-made vehicles can use inertial navigation systems (*41*). These findings are consistent with reports that body-fixed pigeons (*24*), free-standing pigeons (*42*), and hand-held owls (*25*) precisely stabilize their head in the dark based on vestibular and proprioceptive feedback alone.

The innate ability of lovebirds to navigate short distances based on minimal visual information suggests new ways of thinking about crosswind compensation in general. The lovebirds show highly effective compensation compared to the partial compensation of migratory flying animals (Fig. 6A). This difference is understandable given the different constraints on long-range navigation: migratory birds adjust crosswind compensation based on geographic region (*3*, *4*), time of day (*3*), or altitude (*5*), whereas the lovebirds had to fully compensate to reach their goal. The lovebirds' full compensation is consistent with other non-migratory movements, such as bees flying straight—despite high winds—between their nest and their feeding site (*43*). Considering lovebirds are diurnal generalists, their innate ability may apply more generally to birds, nearly all of whom could use proprioception to detect wind over a wingbeat or less (Fig. 6B). In long-range navigation, the dim light heading could be replaced with the celestial, magnetic, and light polarization cues sensed by birds and insects (*6*). Our models therefore provide a basis for understanding how flying animals cope with crosswinds when visual cues are sparse. They may help explain how birds migrate effectively at night and can fly in clouds and fog (*15*, *16*), and how blindfolded gulls can fly stably in crosswinds (*44*). They may also inspire minimal control algorithms that enable aerial robots to maneuver in windy and dark environments as deftly as lovebirds.

## Acknowledgements

We thank P. Neiser and E. Gutierrez for help building the wind generators, E.I Knudsen, L.M. Giocomo, and S.H Collins for helpful discussion and manuscript feedback.

## Funding

Supported by HFSP grant RGP0003/2013, a Stanford Bio-X IIP Seed Grant, the Micro Autonomous Systems and Technology at the Army Research Laboratory – Collaborative Technology Alliance Center grant MCE-16-17-4.3, and NSF CAREER Award 1552419 to D.L.